\pdfoutput=1

\documentclass[11pt,a4paper]{article}
\usepackage{acl}

\usepackage[most]{tcolorbox}

\usepackage{times}
\usepackage{latexsym}
\usepackage{graphicx}
\usepackage{booktabs}
\usepackage[T1]{fontenc}

\usepackage[utf8]{inputenc}

\usepackage{microtype}

\usepackage{inconsolata}

\usepackage{graphicx}

\usepackage{dblfloatfix} 
\usepackage{pifont} 
\usepackage{array} 
\usepackage{makecell} 
\usepackage{ragged2e} 
\usepackage{rotating}
\usepackage{tabularx}
\usepackage{colortbl}
\usepackage{multirow}
\usepackage{enumerate}
\usepackage{longtable}
\usepackage{booktabs}       
\usepackage{url}            
\usepackage{pdflscape}      
\usepackage{float}          
\newcolumntype{P}[1]{>{\centering\arraybackslash}p{#1}}
\newcolumntype{M}[1]{>{\centering\arraybackslash}m{#1}}
\newcolumntype{L}[1]{>{\raggedright\arraybackslash}m{#1}}
\newcolumntype{C}[1]{>{\centering\arraybackslash}m{#1}}
\title{The AI Productivity Index: APEX-v1-extended}

\author{
    \textbf{Bertie Vidgen}$^{1}$ \quad
    \textbf{Abby Fennelly}$^{1}$ \quad
    \textbf{Evan Pinnix}$^{1}$ \quad
    \textbf{Julien Benchek}$^{1}$ \quad \\
    \textbf{Daniyal Khan}$^{1}$ \quad
    \textbf{Zach Richards}$^{1}$ \quad 
    \textbf{Austin Bridges}$^{1}$ \quad
    \textbf{Calix Huang}$^{1}$ \quad \\
    \textbf{Kanishka Sahu}$^{1}$ \quad 
    \textbf{Abhishek Kottamasu}$^{1}$ \quad 
    \textbf{Bo Ma}$^{1}$ \quad 
    \textbf{Ben Hunsberger}$^{1}$ \quad  \\
    \textbf{Isaac Robinson}$^{1}$ \quad
    \textbf{Akul Datta}$^{1}$ \quad
    \textbf{Chirag Mahapatra}$^{1}$ \quad \\
    \textbf{Dominic Barton} \quad 
    \textbf{Cass R. Sunstein}$^{2}$ \quad
    \textbf{Eric Topol}$^{3}$ \quad \\
    \textbf{Brendan Foody}$^{1}$ \quad
    \textbf{Osvald Nitski}$^{1}$\thanks{Email: apex@mercor.com} \\
    $^1$Mercor \quad $^2$Harvard Law School \quad $^3$The Scripps Research Institute 
}

\begin{document}
\pagestyle{plain}
\thispagestyle{plain}

\maketitle
\begin{abstract}
We present an extended version of the AI Productivity Index (\textbf{APEX-v1-extended}), a benchmark for assessing whether frontier models are capable of performing economically valuable tasks in four jobs: investment banking associate, management consultant, big law associate, and primary care physician (MD). This technical report details the extensions to APEX-v1, including an increase in the held-out evaluation set from $n = 50$ to $n = 100$ cases per job ($n = 400$ total) and updates to the grading methodology. We present a new leaderboard, where GPT5 (Thinking = High) remains the top performing model with a score of $67.0\%$.
APEX-v1-extended shows that frontier models still have substantial limitations when performing typical professional tasks.
To support further research, we are open sourcing $n = 25$ non-benchmark example cases per job ($n = 100$ total) along with our evaluation harness.
\end{abstract}

\begin{figure}[t]
\centering
\includegraphics[width=1.05\linewidth]{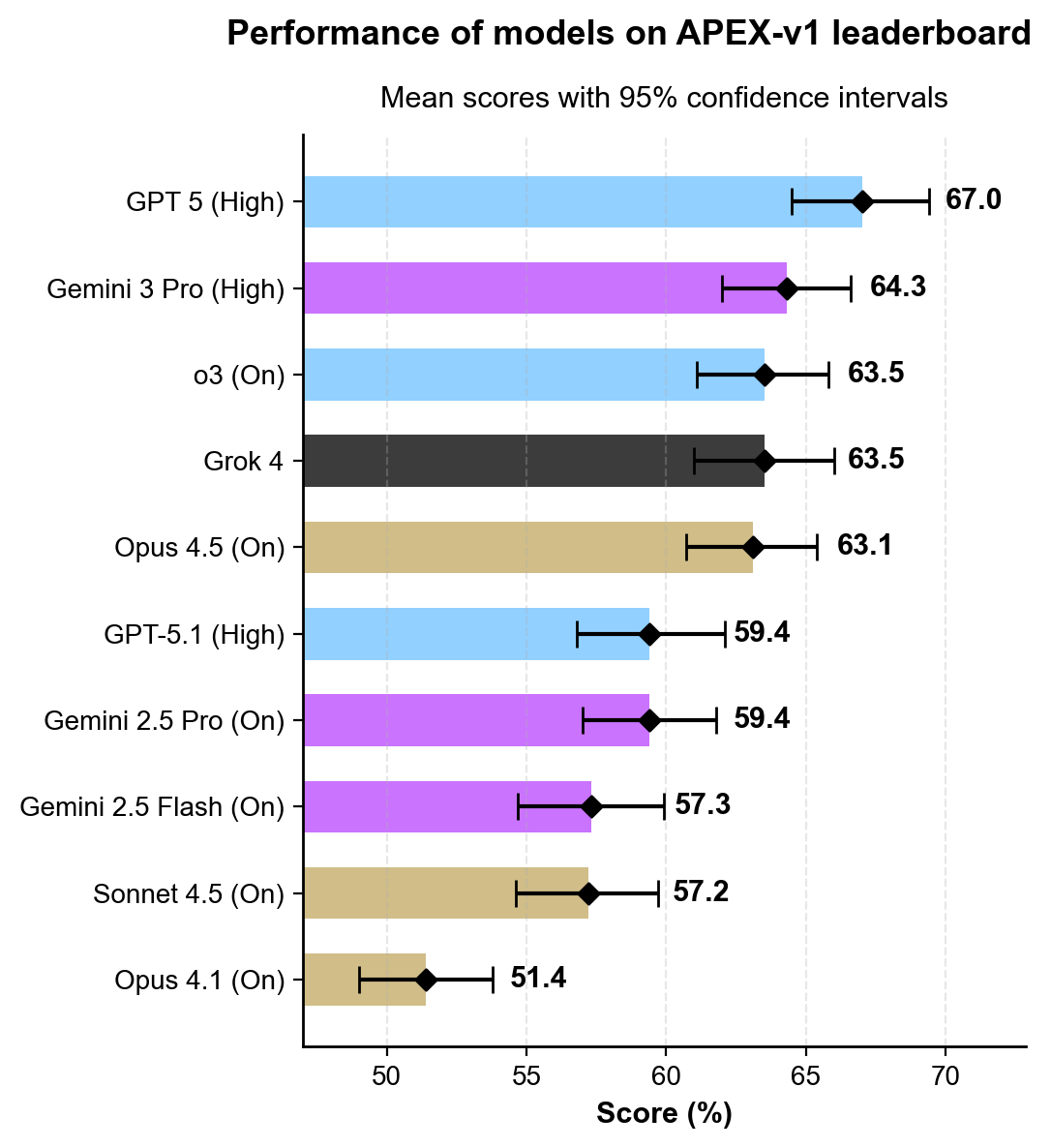}
\caption{Models' mean score on \textbf{APEX-v1-extended}. The labels in parentheses show the ``Thinking'' setting.}
\label{fig:model-overall-mean-score}
\end{figure}

\begin{figure*}[t]
\centering
\includegraphics[width=0.9\linewidth]{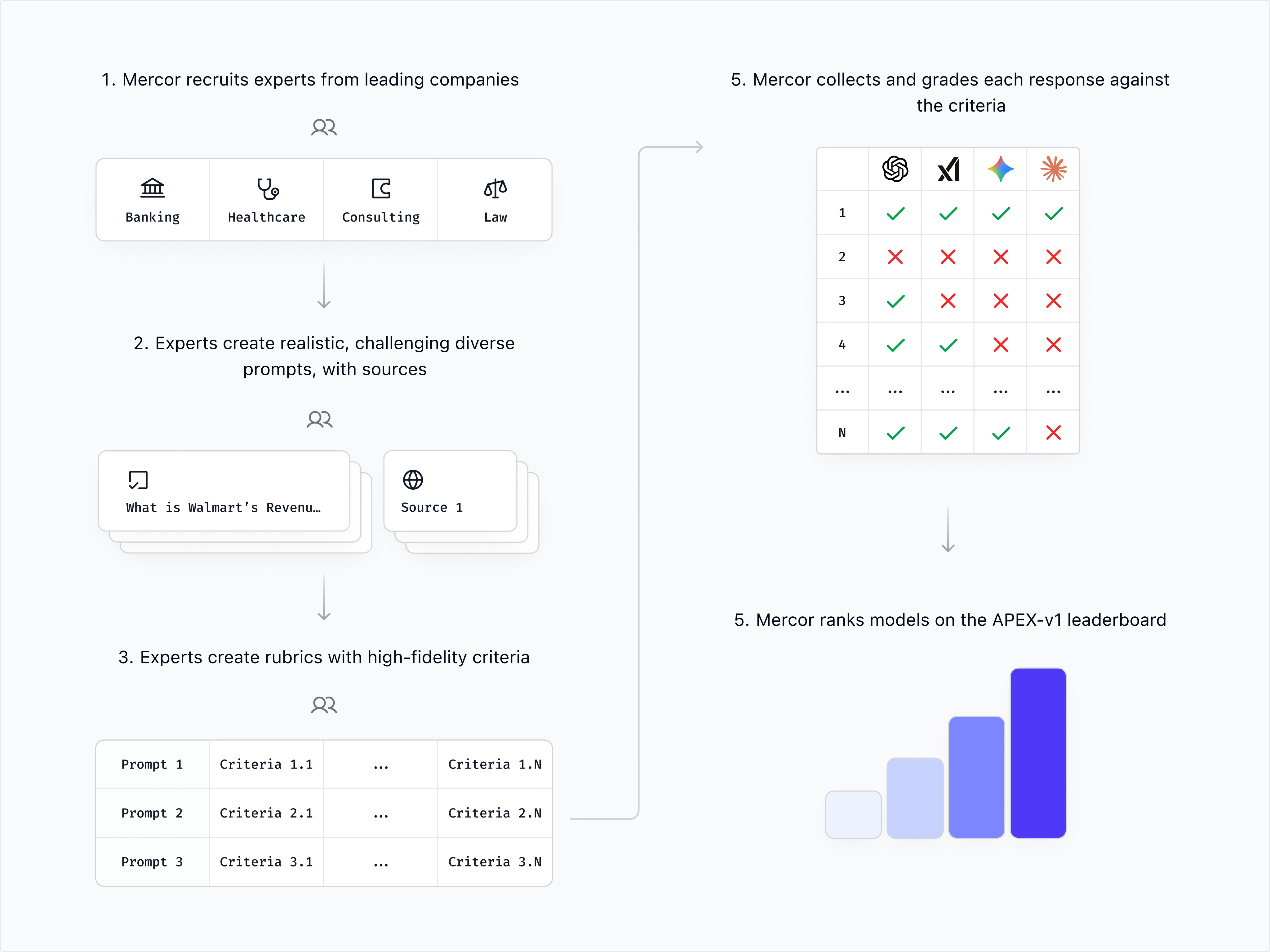}
\caption{Production process for creating \textbf{APEX-v1-extended}.}
\label{fig:model_splash}
\end{figure*}

\setlength{\parindent}{0pt}
\section{Introduction}
AI benchmarks give insight into models' strengths and weaknesses, helping to gauge their capabilities and compare performance. Good benchmarking is essential for effective hill climbing and progress in AI \citep{kiela-etal-2021-dynabench, zhang2024enterprisebenchmarkslargelanguage, schwartz2025realitychecknewevaluation, weidinger2025evaluationsciencegenerativeai}.
However, many benchmarks are scholastic, abstract and narrow, and uncoupled from real-world applications. 
With a few exceptions (such as GDPVal from OpenAI \citep{openai_gdpval} and BigLaw Bench from Harvey \citep{HarveyAI2024BigLawBench}), most benchmarks do not test frontier AI systems for their ability to deliver real world value, such as improving the productivity of human workers \citep{cahn2024AI600B, BarGillSunstein2025, Filippucci2024_MiracleOrMyth, Anthropic_Report_Unknown2025}.
In September 2025 we introduced the AI Productivity Index (APEX-v1.0) to tackle this problem \citet{vidgen2025aiproductivityindexapex}, bridging the gap between what professionals want from AI systems and what benchmarks test for. 
We now introduce \textbf{APEX-v1-extended}. It expands the breadth and robustness of APEX and has an improved grading methodology.
\vspace{1em}

\begin{figure*}[!t]
\centering
\includegraphics[width=0.95\linewidth]{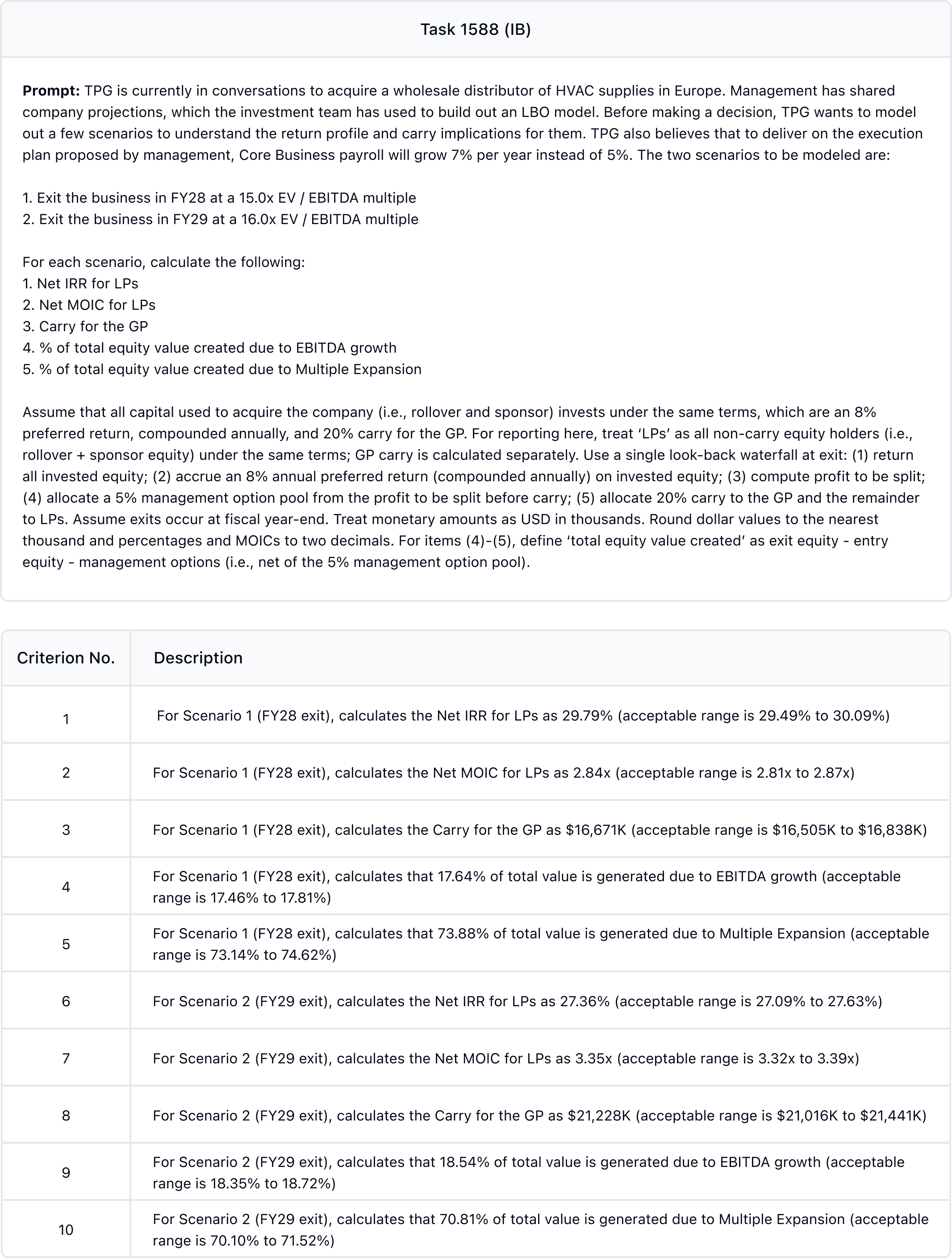}
\caption{Example prompt and rubric from the Investment banking associate job in \textbf{APEX-v1-devset}, with $10$ criteria. The prompt is passed to models with one source, a financial model in an XLSX file.} \label{fig:example-rubric}
\end{figure*}

APEX-v1-extended contains realistic and varied prompts that reflect the everyday work of professionals. They are split across the same four knowledge jobs as the original v1 benchmark: investment banking associate, management consultant, big law associate, and primary care physician (MD). We have increased the size of the hidden heldout set from $n = 50$ to $n = 100$ per job, increasing the overall size of the benchmark from $n = 200 $ to $n = 400$. The new cases have the same design, comprising a  prompt, source documents that the model has to reason over, and a prompt-specific grading rubric. See an example in Figure ~\ref{fig:example-rubric}. The $n = 400$ cases will remain a closed heldout dataset for rigorous evaluation of frontier models. We are also open sourcing a dataset of $n = 25$ cases per job ($n = 100$ total) with a CC-BY license. We refer to this as \textbf{apex-v1-devset}.\footnote{\href{https://huggingface.co/datasets/mercor/APEX-v1-extended}{APEX-v1-extended on Hugging Face}.} Our eval harness is available on Github.\footnote{\href{https://github.com/Mercor-Intelligence/apex-evals}{Mercor on Github}.} 
\vspace{1em}

Changes to grading in APEX-v1-extended include: (1) switching the panel of LM judges to a single LM, Gemini 2.5 Flash, (2) passing the prompts to each model 8 times and taking a mean (rather than 3 times and taking a median) and (3) reporting 95\% confidence intervals with model scores.
We have used this methodology to create a new leaderboard for APEX-v1-extended with results for 10 frontier models.\footnote{\href{https://mercor.com/apex/}{mercor.com/apex}.} 
GPT 5 (Thinking = High) has the highest mean score at $67.0\%$, followed by Gemini 3 Pro (Thinking = High) at $64.3\%$ and Grok 4 at $63.5\%$. 
APEX-v1-extended shows that models struggle with real-world tasks that professionals undertake everyday.
\vspace{1em}

\begin{table*}[htbp]
\small
\centering
\caption{Overview of the \textbf{APEX-v1-extended} dataset. For the heldout evalset ($n = 100$ per job) and the open source data ($n = 25$ per job) we show the mean number of criteria and sources per case and the mean number of tokens in the prompts. Note that the heldout eval set is a recalibrated version of the original set of $n = 50$ with $n = 50$ new cases.}
\label{tab:data-overview}
\begin{tabular}{llcccc}
\toprule
\textbf{Dataset} & \textbf{Job} & \makecell{\textbf{No. of}\\\textbf{cases}} & \makecell{\textbf{No. of }\\\textbf{criteria}}& \makecell{\textbf{No. of}\\\textbf{sources}} & \makecell{\textbf{No. of tokens}\\\textbf{in prompt}} \\

\midrule
Heldout set  & Medicine                 & 100 & 18.39 & 3.77 & 177 \\
Open source  & Medicine                 & 25  & 16.76 & 1.08 & 152 \\
\midrule
Heldout set  & Law                      & 100 & 13.98 & 4.79 & 407 \\
Open source  & Law                      & 25  & 9.88  & 3.32 & 367 \\
\midrule
Heldout set  & Management consulting    & 100 & 14.71 & 3.21 & 409 \\
Open source  & Management consulting    & 25  & 9.68  & 1.16 & 269 \\
\midrule
Heldout set  & Investment banking       & 100 & 12.17 & 3.01 & 439 \\
Open source  & Investment banking       & 25  & 9.28  & 1.60 & 500 \\
\midrule

Heldout set & Mean for all four jobs  & 100 & 14.81 & 3.70  & 358\\
Open source & Mean for all four jobs  & 25  & 11.40 & 1.79  & 322\\
\bottomrule
\end{tabular}
\end{table*}

\section{Dataset overview}
The composition of APEX-v1-extended is shown in Table~\ref{tab:data-overview} and the production process is shown in Figure ~\ref{fig:model_splash}.

\subsection{Selection of experts}
From the Mercor platform we sourced experts with extensive professional experience, and prioritized experts with data labeling experience.
To join the project, experts completed a 30- to 45-minute interview. Experts who demonstrated excellent domain knowledge and strong communication and reasoning skills were paid to complete a 1-2 hour assessment that tested their ability to write prompts and rubrics. If they successfully completed the assessment, they were contracted to work on APEX-v1.0.
Throughout the project we continually checked in with the experts, gave qualitative feedback, and as needed offboarded under-performers.
Expert sourcing and vetting did not change from APEX-v1.0 to APEX-v1-extended.
The number of experts who worked on the benchmark increased from 76 to 137. Mean industry experience remains at 7+ years.
\vspace{1em}

\subsection{Quality control}
Experts created each prompt independently, based on their day-to-day work. The mean time for an expert to execute the prompts in a real-world professional setting is estimated at $2.7$ hours, with a range of $0.5$ to $20$ hours. 
Experts submitted their prompts for review, where they were approved or rejected by reviewers.
For rejections, reviewers could either remove the prompt from production or request specific changes. After the prompt was approved, the contributor created the grading rubric, which was also reviewed by a reviewer. 
Multiple rounds of review help to ensure that cases meet the requirements of the project, are varied in terms of topic, style and model behavior, and are sufficiently challenging. 
We use in-house LM-powered reviewing tools to give experts immediate feedback on their work. Experts are always responsible for the quality of their submissions. 
Finally, to ensure that models give a valid single-turn response, we append a short instruction to each prompt, as given in Appendix~\ref{sec:appendix_prompt_adjustment}.
\vspace{1em}

\subsection{Selection of sources}
For each case, experts found suitable sources or, if they were not freely available, created them. 
Sources with restrictive licenses or access controls, such as being behind a paywall, were explicitly prohibited.
APEX-v1-extended contains pdf, xlsx, txt, docx and csv files, up to a maximum combined length of 100,000 tokens.\footnote{Checked against \href{https://platform.openai.com/tokenizer}{OpenAI's tokenizer for GPT 4o}.}
Sources were parsed using Reducto and appended to the prompt in the context window.\footnote{\href{https://docs.reducto.ai/overview}{reducto.ai}}
The number of sources in APEX-v1-extended is substantially less than in APEX-v1.0, falling from a mean of $5.83$ to $3.70$ in the APEX-v1-extended and $1.79$ in APEX-v1-devset. Experts reported that they could construct realistic prompts by leveraging more content from a single source without the burden of introducing additional sources.
\vspace{1em}

\begin{table*}[htbp]
\small
\centering
\caption{Model performance across the four jobs in \textbf{APEX-v1-extended}. Scores represent the mean percentage of criteria met for tasks within each job. GPT 5 (Thinking = High) is the best performing model overall, and the best performing in Law and Medicine. Gemini 3 Pro (Thinking = High) is the best performing model for Investment banking and Management consulting.}
\label{tab:model-scores}
\begin{tabular}{L{3.2cm} | P{1.5cm} | P{1.3cm} | P{1.3cm} P{1.3cm} P{1.5cm} P{1.3cm}}
\toprule
\textbf{Model} & \textbf{Provider} & \textbf{Overall} & \makecell{\textbf{Invest.}\\\textbf{banking}} & \textbf{Law} & \makecell{\textbf{Mgmt.}\\\textbf{consulting}} & \textbf{Medicine} \\
\midrule
Opus 4.5 (On)         & Anthropic & 63.1\% & 55.2\% & 74.0\% & 58.4\% & 64.6\% \\
Sonnet 4.5 (On)       & Anthropic & 57.2\% & 45.7\% & 72.4\% & 50.7\% & 59.6\% \\
Opus 4.1 (On)         & Anthropic & 51.4\% & 42.0\% & 61.8\% & 49.4\% & 52.1\% \\
Gemini 3 Pro (High)   & Google & 64.3\% & \textbf{63.0\%} & 68.5\% & \textbf{64.0\%} & 61.6\% \\
Gemini 2.5 Pro (On)   & Google & 59.4\% & 54.1\% & 68.5\% & 58.9\% & 55.8\% \\
Gemini 2.5 Flash (On) & Google & 57.3\% & 51.1\% & 68.8\% & 55.6\% & 53.6\% \\
GPT 5.1 (High)        & OpenAI & 59.4\% & 44.3\% & 77.4\% & 51.9\% & 64.0\% \\
GPT 5 (High)          & OpenAI & \underline{\textbf{67.0\%}} & 61.3\% & \textbf{77.9\%} & 63.1\% & \textbf{65.5\%} \\
o3 (On)               & OpenAI & 63.5\% & 57.7\% & 75.5\% & 59.0\% & 61.5\% \\
Grok 4                & xAI & 63.5\% & 59.6\% & 70.2\% & 59.8\% & 64.3\% \\
\bottomrule
\end{tabular}
\end{table*}

\subsection{Rubric creation}
To scalably grade each model response, experts created a rubric of quality criteria \citep{saadfalcon2024lmunitfinegrainedevaluationnatural, arora2025healthbenchevaluatinglargelanguage, starace2025paperbenchevaluatingaisability}. Rubrics decompose the hard-to-measure concept of response ``quality'' into testable discrete components.
Each criterion is an objective, specific, and self-contained statement about the response, phrased as a descriptive claim. They are analogous to unit tests for code, and can be assessed as Pass or Fail by either a judge LM or a human reviewer.
The mean number of criteria per case is $14.81$ in APEX-v1-extended, compared with $11.40$ in APEX-v1-devset. This is a reduction from $29.09$ criteria in APEX-v1.0, due to (1) new cases being added with fewer criteria and (2) excess criteria being removed from the original cases.   
\vspace{1em}

\section{Leaderboard setup}
\subsection{Model selection}
We tested 10 frontier models from Anthropic, Google DeepMind, OpenAI, and xAI on APEX-v1-extended. Responses were collected at the end of November 2025 from the models' respective APIs. Thinking is turned On if available and set to High if configurable. If temperature can be configured, we set it to $0.7$. For each prompt, we collect model responses 8 times. This is an increase from 3 times in APEX-v1.0, allowing us to calculate the mean (rather than median) and confidence intervals.
\vspace{1em}

\subsection{Grading responses with an LM judge}\label{sec:lm_judges_performance}
LM judges are widely used to scalably assess the quality of model responses, supporting high-fidelity evals that can be re-run with no marginal human work \citep{baumann2025largelanguagemodelhacking, gu2025surveyllmasajudge, zhu2025judgelmfinetunedlargelanguage}. LM judging is well-suited for grading rubrics as each criterion is a short self-contained statement. In principle, these statements are easier to judge than overall quality or abstract concepts like ``usefulness'' or ``helpfulness''. We used a single judge in APEX-v1-extended (Gemini 2.5 Flash) rather than the panel of judges for APEX-v1.0. This reduces the eval time, is more transparent, and maintains high performance. The prompt used by the LM judge remains the same, as given in Appendix~\ref{sec:appendix_judge_lm}.
\vspace{1em}


\section{Results}
\subsection{Leaderboard results}
For each model, we compute the mean score from their eight runs on each task. This gives a more reliable measure of model performance. For all 10 models, the mean standard deviation over the eight runs is $9.05$ percentage points, ranging from $8.25$ (Claude Opus 4.5 (Thinking = On)) to $9.5$ (Grok 4).
\vspace{1em}

GPT 5 (Thinking = High) has the highest mean score at $67.0\%$, followed by Gemini 3 Pro (Thinking = High) at $64.3\%$ and Grok 4 at $63.5\%$. Models' mean scores on APEX-v1-extended are shown in Figure~\ref{fig:model-overall-mean-score} with $95\%$ confidence intervals. The differences between models' scores are statistically significant overall using a Friedman omnibus test ($p < 0.000001$). Models' scores vary across jobs. They are lowest for Investment banking, with the highest scoring model achieving $63.0\%$, followed by Management consulting (top score = $64.0\%$), Medicine (top score = $65.5\%$), and, with substantially higher scores, Law (top score = $77.9\%$). Despite jobs' differing difficulty, and the different knowledge and reasoning skills they require, models' rank positions are fairly consistent.
\vspace{1em}


\subsection{Performance given task difficulty}
Because tasks in APEX-v1-extended are realistic they vary in difficulty, with model scores ranging from $0\%$ to $100\%$. We use paired t-tests on the per-task scores for every pair of models ($n = 45$) to directly compare performance. After applying a Bonferroni correction for multiple comparisons, 35 of the 45 pairwise comparisons ($78\%$) remain statistically significant at the adjusted threshold ($p = 0.001$). This is also reflected in the z-scores for each model, which indicate how many standard deviations a model’s performance on that task is above or below the mean performance of all models, as shown in Table~\ref{tab:model_z_scores}. Z-scores are useful for comparing models but are not suitable for a leaderboard because their values change depending on the set of models that are included. With z-scores, the differences between models are much more stark, showing the substantially better performance of GPT 5 (Thinking = High), which scores 0.50 compared to the second best model, Opus 4.5 (Thinking = On) at 0.28. The rank order of the models changes slightly, with Opus 4.5 (Thinking = On) jumping from 5th to 2nd. This indicates that it is particularly strong on the hardest tasks in APEX-v1-extended.
\vspace{1em}

\begin{table}
\small
\centering
\caption{Mean z-scores with 95\% confidence intervals for 10 models on \textbf{APEX-v1-extended}.}
\label{tab:model_z_scores}
\begin{tabular}{l|c|c}
\toprule
\textbf{Model} & \textbf{Z-score} & \textbf{95\% CI} \\
\midrule
GPT 5 (High)          & 0.50 & 0.42 -- 0.58 \\
Opus 4.5 (On)         & 0.28 & 0.20 -- 0.36 \\
Gemini 3 Pro (High)   & 0.20 & 0.10 -- 0.29 \\
o3 (On)               & 0.19 & 0.11 -- 0.27 \\
Grok 4                & 0.15 & 0.05 -- 0.24 \\
GPT-5.1 (High)        & -0.00 & -0.10 -- 0.09 \\
Gemini 2.5 Pro (On)   & -0.10 & -0.18 -- -0.02 \\
Sonnet 4.5 (On)       & -0.18 & -0.27 -- -0.10 \\
Gemini 2.5 Flash (On) & -0.25 & -0.34 -- -0.16 \\
Opus 4.1 (On)         & -0.77 & -0.86 -- -0.68 \\
\bottomrule
\end{tabular}
\end{table}

\subsection{Comparison of APEX-v1-extended and APEX-v1-devset}
We evaluate the same 10 models against the $n = 100$ cases in APEX-v1-devset using the exact same grading methodology (i.e., 8 runs and Gemini 2.5 Flash as a judge). Results are shown in Table~\ref{tab:bench_os_comparison}. GPT 5 (Thinking = High) is the best scoring model on both sets, achieving $65.70\%$ on the APEX-v1-devset versus $67.0\%$ on APEX-v1-extended. Differences in model ranks are minor, with two sets of two models swapped one position, and only one model performs better on APEX-v1-devset than APEX-v1-extended (Gemini 3 Pro (Thinking = High) scoring $1.2\%$ higher). GPT 5.1 (Thinking = High), Gemini 2.5 Flash (Thinking = On) and Sonnet 4.5 (Thinking = On) all score lower on APEX-v1-devset by more than 10 percentage points.

\begin{table}
\small
\centering
\caption{Comparison of model scores and ranks between the \textbf{APEX-v1-extended} leaderboard and the \textbf{APEX-v1-devset}.}
\label{tab:bench_os_comparison}
\begin{tabular}{l|c|c|c}
\toprule
\textbf{Model} &
\begin{tabular}[c]{@{}c@{}}\textbf{Bench }\\\textbf{(rank)}\end{tabular} &
\begin{tabular}[c]{@{}c@{}}\textbf{OS }\\\textbf{(rank)}\end{tabular} &
\textbf{Diff} \\
\midrule
GPT 5 (High)          & 67.0 (\#1) & 65.7 (\#1)  & 1.3 \\
Gemini 3 Pro (High)   & 64.3 (\#2) & 65.5 (\#2)  & -1.2 \\
Grok 4                & 63.5 (\#3) & 61.4 (\#3)  & 2.1 \\
o3 (On)               & 63.5 (\#4) & 61.2 (\#4)  & 2.3 \\
Opus 4.5 (On)         & 63.1 (\#5) & 54.7 (\#5)  & 8.4 \\
GPT-5.1 (High)        & 59.4 (\#6) & \textbf{48.2 (\#7)}  & 11.3 \\
Gemini 2.5 Pro (On)   & 59.4 (\#7) & \textbf{53.4 (\#6)}  & 6.0 \\
Gemini 2.5 Flash (On) & 57.3 (\#8) & \textbf{45.4 (\#9)}  & 11.9 \\
Sonnet 4.5 (On)       & 57.2 (\#9) & \textbf{47.0 (\#8)}  & 10.2 \\
Opus 4.1 (On)         & 51.4 (\#10)& 43.8 (\#10) & 7.6 \\
\bottomrule
\end{tabular}
\end{table}


\section{Acknowledgments}
We are grateful to the expert annotators who contributed to APEX. We thank all our advisors and partners who gave feedback.

\bibliography{anthology,custom}
\bibliographystyle{acl_natbib}

\appendix

\section{LM judge}\label{sec:appendix_judge_lm}
\textbf{Implementation details:}

You are evaluating a model-generated response against a specific criterion. Your task is to determine if the response satisfies this criterion and provide a concise explanation.
\newline \newline
\textbf{Criterion to evaluate:} \texttt{evaluation\_case}
\textbf{Response to evaluate:} \texttt{model\_response}
\newline \newline
\textbf{Instructions:}
\begin{enumerate}
    \item First, analyze the response against the given criterion.
    \item Determine if the response fully satisfies the criterion (\texttt{result = 1}) or not (\texttt{result = 0}).
    \item Provide a concise explanation (maximum 2--3 sentences) that:
    \begin{itemize}
        \item States whether the criterion is met or not,
        \item Points to specific evidence from the response,
        \item Avoids unnecessary details or repetition.
    \end{itemize}
\end{enumerate}

\textbf{Return your evaluation in the following JSON format:}
\begin{verbatim}
    {
    "result": <1 or 0>,
    "reason": "<concise explanation>"
    }
\end{verbatim}

Keep your explanation brief and focus on the key points that justify your result.

\section{Prompt adjustment}\label{sec:appendix_prompt_adjustment}
Many models are trained to ask followup questions or outline options that the user can confirm or reject. To ensure they give a complete response, we append the following text to every prompt.

\subsection*{**Instruction**}
You are an AI assistant that produces final, domain-appropriate deliverables from a given task description and (optionally) attached files. You will be given the following inputs:

\begin{itemize}
    \item Task Domain: {Domain}
    \item Task Prompt: {Prompt}
    \item Attachments:
\end{itemize}

\begin{verbatim}
  ==== Attached files content: ====
  === <File_1> ===
  <File_1_Contents>
  === <File_2> ===
  <File_2_Contents>
  … (repeat as needed)

\end{verbatim}

\textbf{Ground Rules}
\begin{enumerate}
    \item You must not ask follow-up questions. Interpret the prompt as best you can and produce the best complete answer given the information provided.
    \item Use the attachments as primary sources.
    \item Treat each "=== <File\_Name> === …" block as the full content of that file.
\end{enumerate}

All of the source files that you need have been added to the prompt.

\end{document}